

\input phyzzx.tex

%
\catcode`\@=11 
\def\papers{\papersize\headline=\paperheadline\footline=\paperfootline}
\def\papersize{\hsize=40pc \vsize=53pc \hoffset=0pc \voffset=1pc
   \advance\hoffset by\HOFFSET \advance\voffset by\VOFFSET
   \pagebottomfiller=0pc
   \skip\footins=\bigskipamount \normalspace }
\catcode`\@=12 
\papers
\def\to{\rightarrow}

\tolerance=500000
\overfullrule=0pt

\pubnum={PUPT-1440}

\date={Dec 1993}
\pubtype={}
\titlepage
\title{An Exact Solution to O(26) Sigma Model
\break
Coupled to 2-D Quantum Gravity
}
\author{{
Weidong Zhao}\foot{
work supported by NSERC postdoctoral fellowship\nextline
e-mail: wdz@puhep1.princeton.edu
}}
\address{\it Joseph Henry Laboratories\break
Princeton University\break
Princeton, NJ
08544, USA}

\vskip 3.mm
\abstract{By a mapping to the  bosonic string theory,  we present an exact
solution to the O(26) sigma model coupled to
2-D quantum gravity. In particular, we obtain the exact
gravitational dressing to
the various matter operators classified by the
 irreducible representations of O(26).
We also derive the exact form of the gravitationally modified beta function
for the original coupling constant $e^2$. The relation between our
exact solution and the asymptotic solution given in ref[3] is discussed
in various aspects.

}

\endpage
\pagenumber=1

 \def\PL #1 #2 #3 {Phys.~Lett.~{\bf #1} (#2) #3}
 \def\NP #1 #2 #3 {Nucl.~Phys.~{\bf #1} (#2) #3}
 \def\PR #1 #2 #3 {Phys.~Rev.~{\bf #1} (#2) #3}
 \def\PRL #1 #2 #3 {Phys.~Rev.~Lett.~{\bf #1} (#2) #3}
 \def\CMP #1 #2 #3 {Comm.~Math.~Phys.~{\bf #1} (#2) #3}
 \def\IJMP #1 #2 #3 {Int.~J.~Mod.~Phys.~{\bf #1} (#2) #3}
 \def\JETP #1 #2 #3 {Sov.~Phys.~JETP.~{\bf #1} (#2) #3}
 \def\PRS #1 #2 #3 {Proc.~Roy.~Soc.~{\bf #1} (#2) #3}
 \def\IM #1 #2 #3 {Inv.~Math.~{\bf #1} (#2) #3}
 \def\JFA #1 #2 #3 {J.~Funkt.~Anal.~{\bf #1} (#2) #3}
 \def\LMP #1 #2 #3 {Lett.~Math.~Phys.~{\bf #1} (#2) #3}
 \def\IJMP #1 #2 #3 {Int.~J.~Mod.~Phys.~{\bf #1} (#2) #3}
 \def\FAA #1 #2 #3 {Funct.~Anal.~Appl.~{\bf #1} (#2) #3}
 \def\AP #1 #2 #3 {Ann.~Phys.~{\bf #1} (#2) #3}
\def\MPL #1 #2 #3 {Mord.~Phys.~Lett. ~{\bf #1} (#2) #3}

  \def\ins{\int d\sigma^2}
\def\al{\alpha}
\def\pd{\partial}
\def\g0{\hat g}
\def\al{\alpha}
\def\be{\beta}

Consider the following two dimensional O(N) nonlinear
sigma model\REF\POLY{See, for example, A.Polyakov, {\it Gauge Fields
and Strings},
Harwood Academic Publishers (1987)} [\POLY]
$$I={1\over e^2}\ins \pd n^i\pd n_i. \eqno(1)$$
This model is well known to exhibit such features  as
the  asymptotic freedom at the UV limit, and dimensional transmutation
leading to
massive modes as low-lying excitations, etc. An easy way to
visualize   these properties is  through the large N expansion, in which
 the mass of
the vector-like excitation is related to the bare coupling constant by
$m^2=\Lambda^2 \exp(-{4\pi\over Ne^2})$.  An exact solution to this model
is also available  by the means of Bethe Ansatz\REF\ZZ{A.B.Zamolodchikov and
Al.B. Zamolodchikov, \AP 120 1979 253 } [\ZZ].

We may ask what would happen if this model is coupled to the 2-D
  gravity. This is
a problem of the gravitational dressing to a
renormalizable but not conformally invariant 2-D field theory. This problem
has recently attracted some attention, see \REF\KKP{I.Kogan, I.Klebanov and
A.Polyakov, Princeton preprint PUPT-1421, September 1993, hep-th/9309106}
\REF\AMB{J.Ambjorn and Kazuo Ghoroku, preprint NBI-HE-93-63, Nov. 1993;
Y.Tanii, S.Kojima and N.Sakai, preprint Imperial/TP/93-94/10; TIT/HEP-244,
 Nov. 1993}
Ref.[\KKP,\AMB]. In particular,
a modification to the beta functions in the presence of
 gravity has been
derived to the one-loop level[\KKP]. The modification, suspected to be
universal, is verified successfully
in several cases including  the gravitational dressing
to the O(N) sigma model (1). It is important to extend
these results
to higher loops, and if possible, to obtain  exact solutions to this problem.

In analogy to the 2-D conformal matters coupled to  gravity,
there are the light-cone and conformal gauge approaches
 available  to this problem\REF\KPZ{V.Knizhnik, A.Polyakov
and A.Zamolodchikov, \MPL A3 1988 819}\REF\DDK{F.David, \MPL A3 1988 1651;
J.Distler and H.Kawai, \NP B321 (1989) 509} [\KPZ,\DDK].
 The conformal gauge approach,
known as the DDK approach in the case of conformal
matter coupling to gravity, requires to introduce a fiducial world-sheet
 metric
$\g0$, with respect to which the gauge fixed theory is  exactly
conformal invariant.
  As usual, the Liouville
mode $\phi$ serves as the only dynamic degree of freedom left
from gravity, and
its coupling to the matter is encoded in  a nonlinear sigma model with
both $\phi$ and the matter fields as coordinates. The  conformal
invariance can be implemented by requiring the beta functions of this sigma
model
to
vanish exactly. This procedure is reduced  to the DDK approach
 when the bare matter is conformally invariant.
 When  the bare matter is not conformally invariant, one must in principle
solve
 the
equation
of the exact
beta functions. Up to the one-loop level, these beta functions read
\REF\CAL{C.Callen, E.Martinec, M.Perry and D.Friedan, \NP B262 1985 593 }[\CAL]
$$\eqalign{\be^G_{ij}&=R_{ij}-\nabla_i\nabla_j\Phi \cr
\beta^{\Phi}&={26-N\over 3}-\nabla^2\Phi-(\nabla \Phi)^2 }\eqno(2)$$
 where $G_{ij}$ is the target space metric and $\Phi$ is the dilaton
background. In principle, the beta functions at higher orders can
 be derived by the
means of background expansion. Unfortunately, this
 has
not been worked out beyond the fourth order, therefore at this stage
a solution to
the exact beta function is generally impossible.  Furthermore,
even the ones already being
derived, such as (2),
have  been too complicated to solve exactly. In general, this is a
very hard problem.

The situation may be improved if some symmetries are involved. For instance,
the model (1) possesses an O(N) symmetry. It is reasonable to
expect this symmetry  to survive  the gravitational dressing. In other words,
we assume that the  nonlinear
sigma model  have a background  that respects  the symmetry O(N),
under
which $n^i$ and $\phi$ transform as vector and scalar, respectively.
It is easy to observe that the only possible background satisfying
this condition is of the following
form
$$I=\ins \sqrt{\g0} \g0^{\al\be} (b(\phi)\pd_{\al}\phi\pd_{\be}\phi+
a^2(\phi)\pd_{\al}n^i\pd_{\be}n^i)+ \ins \sqrt{\g0} R(\g0)\Phi(\phi).
\eqno(3)$$
Now the general background dependence has  been reduced to through
three unknown
functions
$b, a$ and $\Phi$. Owing to the unitary normality of the vector $n$, these
functions depend on
 $\phi$ only, and
terms proportional to $n^i\pd_{\alpha}n^i$ do not appear in (3).

The beta function equation for (3) is a set of differential
equations for $b,a$ and $\Phi$. Up to the one-loop level, they can be obtained
by substituting the background of (3) into (2),  resulting
$$ (N-2) + {aa'b'\over 2b^2}- {aa''\over b}-(N-2){a'^2
\over b}
={aa'\over b}\Phi' \eqno(4a)$$
$$(N-1)(-{a''\over a}+{a'b'\over 2ab})=\Phi''-{b'
\Phi'\over 2b}
 \eqno(4b)$$
$${\Phi''\over b}+(N-1){a'\Phi'\over ab}-{b'
\Phi'\over 2b^2}+(\Phi')^2=
{26-N \over 3}  \eqno(4c)$$
Note that letting $b=1$  reduces these equations to the ones
obtained in Ref.[\KKP],  in which an asymptotic solution for large $\phi$ was
found
$$\eqalign{b&=1, \cr
\Phi &=Q\phi + {\cal O} (\log \phi),  \cr
a^2&=2{N-2\over Q}\phi +{\cal O} (\log \phi), }\eqno(5)$$
where $Q=\sqrt{{25-c\over 3}}$ and $c=N-1$. Also, the leading asymptotic form
for the cosmological constant operator was identified as
$$V^{(0)}\sim e^{-\al \phi}  \eqno(6)$$
with $\al ={\sqrt{25-c}-\sqrt{1-c}\over \sqrt{12}} $.

Although  (4) is much simpler than the original
equation (2),  to find an exact solution to this equation is still
 a hard task.
Surprisingly,
 in the case of  $N=26$, in which even the
asymptotic form (5) breaks down because of $Q=0$,  an exact solution to (4)
exists.
This solution, satisfying the one-loop equation (4), turns out
to be also an exact solution to the beta function equation {\it to all orders}.
In other words, the background configuration $b,a,\Phi$ given by this
solution will correspond to a conformal field theory with $c=0$ (with ghost
contribution included). Our
purpose
in this paper is to show how these happen. The key observation
here will be that the model (3) can be mapped
to a trivial $D=26$ bosonic string.  As a result,
the complete set of physical operators may be constructed from its counterpart
in the D=26 string theory.
We
shall also
derive the exact expressions for the gravitational dressing to  $n^i$ and
various composite operators of $n^i$. Finally,  we  identify the
 gravitational dressing to the identity as the cosmological
constant operator, and obtain the exact gravitationally dressed
RG equation in the spirit of Ref.[\KKP].

We begin by providing the following solution to (4)
$$b=1; \  \  a^2=\phi^2; \ \  \Phi=0    \eqno(7)$$
It is trivial to verify that (7) indeed satisfies the one-loop equation (4).
To see that it is also an exact solution
beyond the one-loop
level, one may want to write down the beta function equations at two-
and three loops and check the solution explicitly. However, the best
way to convince oneself for
this fact is to consider a D=26 critical string in the
flat target space, whose action is familiar
$$I={1\over 2}\int\sqrt{g}g^{\al\be}\pd_{\al}x^i\pd_{\be}x^i.  \eqno(8)$$
Together with the ghost contribution, this is a conformal field theory with
$c=0$. This fact implies that the beta function equation, even though
its  exact expression is not known, must be  satisfied
by the trivial background $\Phi=0$ and  $R_{ijkl}=0$ of the theory (8).

 Now consider the following coordinate transformation
$$x^i=\phi n^i, \eqno(9)$$
where $  n^in_i=1$ and $\phi$ is the norm of the 26-component vector $x^i$.
Substituting (9) into
(8), we arrive at the following action
$$I={1\over 2}\int\sqrt{g}g^{\al\be}(\pd_{\al}\phi\pd_{\be}\phi + \phi^2
\pd_{\al}n^i\pd_{\be}n_i) \eqno(10)$$
This is exactly the same action as (3), provided the background (7)
 is used.  Since (10) is derived from (8), therefore it is also a conformal
 field theory
with $c=0$. In particular, we still have $\Phi=R_{ijkl}=0$, as (9) is
just a coordinate
transformation and should not change the zero value of a vanishing tensor.
We conclude that the background in (10) must also satisfy the exact
 beta function
equation, even though its exact expression in terms of $b,a,\Phi$ is not given.

One might wonder whether this is the only solution to the problem. At
 this
moment we are unable to provide a definite answer to this question.
A  more appropriate
question to ask might be whether
this solution produces  behaviours more or less
anticipated from  a theory of this type. First of all, let us
compare
the solution (7) with the asymptotic form (5).
The behaviour of $\Phi$  in both solutions agrees, since
the asymptotics for $\Phi$ actually becomes exact as $Q=0$ and no
logarithmic correction is received.
As to $a^2$, the coefficient ${2N-2\over Q}$ for $\phi$ is infinite.
If we interpret
this infinite coefficient
as the other $\phi$ in the exact solution (7) (since $\phi\to\infty$), then
(7)   does not contradict to (5). Of course, behind this naive
interpretation is the statement that the linear asymptotic expansion (5)
breaks
 down at $N=26$.
Later, we shall see another piece of evidence from
the  cosmological constant operator,
which exhibits an asymptotic behaviour
 compatible to the
leading asymptotic behaviour predicted by (6) for $N=26$.

The connection (9) between the gravitationally dressed
O(26) model and the trivial bosonic string is very useful in
computation. In fact, one may be able to
express the various quantities in the former model in
terms of quantities in the
latter model, and then taking advantages of the well known results of
 the bosonic string. For example, the correlation
function $<\phi^2(z)\phi^2(0)>$ may be
easily evaluated as\footnote{\dag}{When performing the path integral, there
is a difference between (10) and (3). In (3), the integration domain for
$\phi$ is between $-\infty$ and $\infty$, whereas in (10) it is restricted to
positive values. This may be resolved by noticing that the action (10) is
symmetric under $\phi\to -\phi$. It is thus trivial to extend the intergration
 domain to the negative values. Similar extensions of $\phi$ in the
physical operators obtained from (10) may also be performed by a replacement
of $\phi$ by $|\phi|$.}
$$<\phi^2(z)\phi^2(0)>=
\sum_{i,j=1}^{26}<x^i(z)x^i(z)x^j(0)x^j(0)>=52(log|z|)^2  \eqno(11)$$
Thus, $\phi^2$ is not a scaling operator. One may want to look for
functions of $\phi$ that  are  scaling operators. Actually, in
such a theory of DDK type, the truly meaningful scaling operators are
those primary operators with conformal dimension (1,1). In the
bosonic string theory, these physical
operators have  been constructed in terms of vertex operators
a long time ago. If we
make
use of the connection (9), we can directly write down the physical operators
for the theory (10). For example, the familiar
tachyon vertex operator in the model (8)
 corresponds to the following operator in the model (10)
$$ V(k, z)=e^{ik^ix_i(z)}=e^{ik^in_i(z)|\phi|(z)}, \  (k^ik_i=2) \eqno(12)$$
In fact, one can take any physical vertex operator in the bosonic string
and simply
replace the coordinates $x^i$ by $|\phi| n^i$,  obtaining a physical
operator for the  model (10). By this way one
can in principle construct the complete set of physical operators for the
gravitationally dressed O(26) model.

The physical operators constructed this way are typically unfactorized
between $\phi$ and $n^i$, which is not convenient for the purpose of
 understanding the
 problem of
gravitational dressing. This undesired feature is a consequence that
 vertex operators such
as $V(k,z)$ are essentially  in the  momentum representation, and the
coordinate
transformation
(9) only preserves the angular momentum manifestly. In order to conform to
the
relation (9), one is  motivated to consider the vertex operators in
the angular momentum
 representation. For the tachyon vertex operator $V(k,z)$, this construction
can be performed as follows. We first observe that any linear combination
of physical operators is still a physical operator, because the
condition for being a physical operator,
$$ L_0 V=\bar{L}_0V=V; \ \ L_n V=\bar{L}_nV=0, \  \ n=1,2,.... \eqno(13)$$
is a linear condition. Consequently, the operator
$\int d^{26}k f(k) \delta(k^2-2)V(k,z)$ for any $f(k)$ is still a primary
operator with conformal dimension (1,1). In fact, we have
$$\eqalign{
&\int d^{26}k \int d^{26}qf(k)f(q)\delta(k^2-2)
\delta(q^2-2)<V(k,z)V(q,0)>  \cr
&=\int d^{26}k \int d^{26}qf(k)f(q)\delta(k^2-2)
\delta(q^2-2)\delta^{26}(k+q)z^{k\cdot q}\bar{z}^{k\cdot q}\cr
&\prop z^{-2}\bar{z}^{-2}.}\eqno(14)$$
Let us now
 define the following tensor operator
 $$V^{(n)}_{i_1...i_n}=(-i)^n\int {d^{26}k\over (2\pi)^{13}} \delta
(k^2-2)k_{i_1}...k_{i_n}e^{ik^in_i|\phi|}
  \eqno(15)$$
For $n=0$, $V^{(0)}$ is a scalar, and must be a function of $\phi$ only.
In fact, the integration can be performed straightforwardly, and the result is
$$V^{(0)}(\phi)=32|\phi|^{-12}J_{12}(\sqrt{2}|\phi|)   \eqno(16) $$
This seems to be the only primary operator with conformal weight (1,1),
which at the same time  is  a local function of $\phi$ only. We  thus
interpret it
as the cosmological
constant operator, corresponding to the gravitational dressing to the identity.
Indeed, in the large $\phi$ limit, $V^{(0)}$ has the following
expansion
$$\eqalign{V^{(0)}(\phi)=32\cdot 2^{1\over 4}|\pi|^{-{1\over 2}}
|\phi|^{-{25\over 2}}
&[\cos
(\sqrt{2}|\phi|-{\pi\over 4})\sum_{m=0}^{\infty}{(-1)^m(12,2m)\over
 (2\sqrt{2}|\phi|)^{2m}}\cr
&-\sin(\sqrt{2}|\phi|-{\pi\over 4})\sum_{m=0}^{\infty}{(-1)^m(12,2m+1)\over
(2\sqrt{2}|\phi|)^{2m+1}}]} \eqno(17)$$
where $(p,m)={\Gamma({1\over 2}+p+m)\over m!\Gamma({1\over 2}+p-m)}$. For large
$\phi$, we retain only the leading term in (17)
$$V^{(0)}(\phi)\sim \exp^{-i
(\sqrt{2}|\phi| -{\pi\over 4})-{25\over 2}log|\phi|}  + c.c.
\eqno(18)$$
This result
agrees with the eqn. (6) (for N=26, $\al=-i\sqrt{2}$)
up to a log$\phi$ correction. \footnote{\dag}{For $c>1$, both the two roots
of $\alpha$ should be retained for a real result, and the usual
 arguments of analytic continuation, which results in
only $\al_- $ as $c\to -\infty$, does not
apply because of the c=1 barrier.} This agreement not only supports
the interpretation of $V^{(0)}$ as the cosmological constant operator, but
also
serve as  further evidence for (7) to be the right solution with
expected asymptotic behaviours.

{}From mathematical point of view, above comparison  between (18) and (6)
makes only formal sense,
since $\alpha$ now is purely imaginary and the subleading logarithmic term
cannot be neglected. What is  more serious is that with this imaginary
$\alpha$ the operator $V^{(0)}(\phi)$ is not positively definite.
This difficulty is well known in
any $c>1$ theory, and in the present case we   offer no better resolution
to
the problem. It is clear that a sensible
prescription (perhaps  some kind of analytic continuation)
is needed in order
for these vertices to make physical sense.
For the time being, at least formally, we assume $V^{(0)}$
to play the role of the cosmological constant operator.

With $V^{(0)}$ being interpreted as the cosmological constant operator, we can
obtain
 the exact expression for the gravitationally dressed
beta function for the coupling constant
$e^2={1\over a^2}$. Following the Ref.[\KKP],
we let $\Lambda^2=V^{(0)}(\phi)$. Since  $e={1\over \phi}$, we have
$$\eqalign{{de^2\over d \log \Lambda}&=\beta(e^2)=
{d\phi\over d\log\Lambda}{de^2\over d\phi} \cr
&=2\sqrt{2}e^{3\over 2}{J_{12}(\sqrt{2\over e^2})\over
 J_{13}(\sqrt{2\over e^2})}}\eqno(19)$$
This beta function is nonperturbative in nature. Actually, the function
$J_{\nu}(x)$ for large $x$ has the following asymptotic expansion
$$J_{\nu}(x)=\sqrt{2\over \pi x}\left(\cos(x-{\nu \pi\over 2}-{\pi\over 4})
\sum_{m=0}^{\infty}{(-1)^m(\nu,2m)\over
 (2x)^{2m}}-\sin(x-{\nu\pi\over 2}-{\pi\over 4})\sum_{m=0}^{\infty}
{(-1)^m(\nu,2m+1)\over
(2x)^{2m+1}}\right) \eqno(20)$$
For small $e$, one can perturbatively
expand (19)  using (20) for
$\nu=12$ and $\nu=13$. The
occurrence of $\sin(\sqrt{2\over e^2})$ and $\cos(\sqrt{2\over e^2})$
in this perturbative expansion is characteristic of the
nonperturbative nature of the beta function. This may be (formally)
compared with
the usual pattern of instanton effect $e^{-{c\over e^2}}$ if we ignore
the difference between an imaginary $c$ and a real $c$.
It would be very interesting to understand the origin of this instanton-like
behaviour from the field theoretic point of view.

Finally, let us turn to the $V^{(n)}$ for general $n$. We will see that
these $V^{(n)}$ shall produce the gravitational dressing for matter
tensors of higher ranks.  In fact, $V^{(n)}$ in
(15) can  be computed for any $n$, resulting
$$V^{(n)}_{j_1...j_n}= 2^{22+n\over 2}\sum_{p=0}^{[{n\over 2}]}
\sum_{1\leq i_1 < ...< i_{2p}\leq n}(-1)^pF^{n,2p}
(\sqrt{2}|\phi|)n_{j_1}..\hat n_{j_{i_1}}.. \hat n_{j_{i_{2p}}}..n_{j_n}
P_{j_{i_1}...j_{i_{2p}}}  \eqno(21)$$
where
$$P_{j_1...j_{2p}}=\delta_{j_1j_2}\delta_{j_3j_4}...
\delta_{j_{2p-1}j_{2p}} + {\rm symmetric\  permutations}\eqno(22)$$
is the total symmetric tensor of rank $2p$, and
$F^{n,2p}(x)=x^{-12-p}J_{12+n-p}(x)$.
As usual, a hat in (21) represents  a missing index. Obviously, for $n=0$, (21)
agrees with (16).

Let us see some examples now.
For $n=1$, we have
$$V^{(1)}_{j}=2^{23\over 2}F^{1,0}(\sqrt{2}|\phi| )n_{j}=
2^{11\over 2}\phi^{-12}J_{13}(\sqrt{2}|\phi|)n_{\mu} \eqno(23)$$
Thus, $F^{1,0}(\sqrt{2}|\phi|)$ can be naturally interpreted as
the gravitational dressing
for the matter operator $n_{j}$. For $n=2$, the eqn.(21) becomes
$$V^{(2)}_{ij}=2^{12}(F^{2,0}(\sqrt{2}|\phi|)n_{i}n_{j}-F^{2,2}
(\sqrt{2}|\phi|)\delta_{ij}) \eqno(24)$$
In this expression $\phi$ and $n$ are  not factorized. The reason is
that $V^{(2)}_{ij}$ is not in an irreducible representation
of O(26). In fact, if we consider the traceless part of $V^{(2)}_{ij}$
$$-\int {d^{26}k\over (2\pi)^{13}} \delta
(k^2-2)(k_{i}k_{j}-{1\over 13}\delta_{ij})e^{ik^in_i|\phi|}
=V^{(2)}_{ij}+{1\over 13}V^{(0)}\delta_{ij} \eqno(25)$$
It is easy to prove, using the identity ${2n\over x}J_n(x)-J_{n-1}(x)
=J_{n+1}(x)  $,
 that this traceless tensor
is equal to
$$ 2^{12}F^{2,0}(\sqrt{2}|\phi|)(n_{i}n_{j}-{1\over 26}\delta_{ij}) \eqno(26)$$
Now  $F^{2,0}(\sqrt{2}|\phi|)$ can be interpreted as the gravitational
dressing of the rank 2 traceless tensor
$n_{i}n_{j}-{1\over 26}\delta_{ij}$.

The lesson we learn from the above example of $n=2$ is that, it is the
irreducible representation of tensor $n_i$ that receives the multiplicative
gravitational dressing. This is natural  because the theory has
 the symmetry of O(26), so its spectrum and Liouville dressing should be
classified by the irreducible representations of O(26). From this point of
view, we can determine the gravitational dressings for tensors of higher ranks
as follows.

Since all the tensors of higher ranks are
constructed from the vector $n^i$, they must be totally symmetric.
For O(N) group, the  only totally symmetric
tensors  corresponding
to  irreducible representations  are  the traceless tensors\REF\GROUP{See, for
example, M.Hamermesh, {\it Group Theory and Its Application
to Physical Problems}, Addison Wesley Inc., 1962}[\GROUP].
At any given rank $n$,  such a tensor can be constructed
from the vector
$n^i$ as
$$T_{i_1...i_n}(n)=\sum_{p=0}^{[{n\over 2}]}
\sum_{1\leq j_1 < ...< j_{2p}\leq n} {(-1)^p\over Q(n,p)}
n_{i_1}..\hat{n}_{i_{j_1}}..\hat{n}_{i_{j_{2p}}}..n_{i_n}
P_{i_{j_1}...i_{j_{2p}}}   \eqno(27)$$
where $Q(n,p)=\prod_{q=0}^{p-1}(22+2n-2q)$ for $p>0$, and $Q(n,0)=1$. One
may verify that (27) with such a choice of coefficients indeed satisfies
the traceless condition.
Likewise, for the irreducible traceless tensor constructed from $k_i$,
we have
$$S_{i_1...i_n}(k)=\sum_{p=0}^{[{n\over 2}]}
\sum_{1\leq j_1 < ...< j_{2p}\leq n} {(-1)^p2^p\over Q(n,p)}
k_{i_1}..\hat{k}_{i_{j_1}}..\hat{k}_{i_{j_{2p}}}..k_{i_n}
P_{i_{j_1}...i_{j_{2p}}}  \eqno(28)$$
where the difference  of the coefficients between (27) and (28) is
 due to the difference of the norms of the vectors $k^i$ and $n^i$.
Now, both $S$ and $T$
are symmetric traceless tensors. According to the  arguments
presented above, we conjecture the following relation
$$(-i)^n\int {d^{26}k\over (2\pi)^{13}} \delta
(k^2-2)S_{i_1...i_n}(k)e^{ik^in_i\phi}=2^{22+n\over 2}F^{n,0}
(\sqrt{2}|\phi|)T_{i_1...i_n}(n). \eqno(29)$$
If this relation is held true, $ F^{n,0}
(\sqrt{2}|\phi|)$ in (29) can be  interpreted as the gravitational dressing
 to the
irreducible representation $T_{i_1...i_n}$ of rank $n$.
We have explicitly verified (29) up to $n=4$. However, so far
 we have not been able to give an analytic proof for it.

In conclusion, we have  presented an exact solution to the
O(26) sigma model coupled to gravity, by mapping the model to a trivial
bosonic string theory. As a result, we obtained the exact gravitational
dressing to a variety of matter fields which form irreducible representations
of O(26). Also, we obtained an exact beta function for the coupling constant
$e^2$. This beta function is nonperturbative in nature, and therefore cannot
 be compared directly to  the perturbative results discussed in [\KKP]. In
fact,
at $N=26$, we find $k=-1$, which means the prefactor ${k+2\over k+1}$ in [\KKP]
diverges. It might be useful to conduct a $N=26+\epsilon$ analysis to clarify
the relationship between the perturbative results and our results.

There are still many questions needed to be addressed. Among them, how to
deal with the imaginary $\al$ is certainly of primary importance.
Unfortunately, a solution to this notorious problem seems still out of our
reach.
Another perhaps easier problem is to understand the roles of the
other physical
operators corresponding to the vertex operators of higher mass levels.
They appear to be related to the gravitational dressing  to the matter
operators involving world-sheet derivatives. At this moment, the picture
in this aspect is still unclear.

\ack {It is my pleasure to thank A.Polyakov for discussion.}

\refout

\bye